\documentclass[twocolumn,preprintnumbers,amssymb,amsmath,9pt,superscriptaddress]{revtex4}[10pt]
\usepackage{graphicx,psfrag}
\usepackage{dcolumn}
\usepackage{bm}
\begin{document} 
\input epsf.tex

\newcommand{\beq}{\begin{eqnarray}}
\newcommand{\eeq}{\end{eqnarray}}
\newcommand{\nn}{\nonumber}
\def\ltap{\ \raise.3ex\hbox{$<$\kern-.75em\lower1ex\hbox{$\sim$}}\ }
\def\gtap{\ \raise.3ex\hbox{$>$\kern-.75em\lower1ex\hbox{$\sim$}}\ }
\def\CO{{\cal O}}
\def\CL{{\cal L}}
\def\CM{{\cal M}}
\def\tr{{\rm\ Tr}}
\def\CO{{\cal O}}
\def\CL{{\cal L}}
\def\CM{{\cal M}}
\def\N{{\cal N}}
\def\mpl{M_{\rm Pl}}
\newcommand{\bel}[1]{\be\label{#1}}
\def\al{\alpha}
\def\bt{\beta}
\def\eps{\epsilon}
\def\eg{{\it e.g.}}
\def\ie{{\it i.e.}}
\def\mn{{\mu\nu}}
\newcommand{\rep}[1]{{\bf #1}}
\def\be{\begin{equation}}
\def\ee{\end{equation}}
\def\bea{\begin{eqnarray}}
\def\R{{\mathbb R}}
\def\eea{\end{eqnarray}}
\newcommand{\ra}{\rightarrow}
\def\coeff#1#2{{\textstyle {\frac {#1}{#2}}}}
\def\atop#1#2{\genfrac{}{}{0pt}{}{#1}{#2}}
\def\half{\coeff 12}
\def\l{\ell}
\def\C{{\cal C}}
\def\N{{\cal N}}
\def\T{{\cal T}}
\def\P{{\cal P}}
\def\I{{\cal I}}
\def\G{{\cal G}}
\def\None{\N\,{=}\,1}
\def\Nfour{\N\,{=}\,4}
\def\R{{\mathbb R}}
\def\Rep{{\cal R}}
\def\tr{{\rm tr}}
\def\Re{{\rm Re}\,}
\def\x{\mathbf x}
\def\r{\mathbf r}
\def\t{\mathbf t}
\def\Nc{N_{\rm c}}
\def\Lc{L_{\rm c}}
\def\Nc{N}
\def\S{{\mathbb S}}
\def\Nf{N_{\rm f}}
\def\Z{{\mathbb Z}}
\def\B{{\mathbb B}}
\def\USp{{\it USp}}
\def\Dslash{{\rlap{\raise 1pt \hbox{$\>/$}}D}}
\def\O{{\cal O}}
\def\Otilde{\widetilde \O}
\def\Veff{V_{\rm eff}}
\def\adj{\mathrm {adj}}
\def\Im{\mathrm {Im}}

\title{AdS/CFT and large-$\mathbf N$ volume independence}
\author{Erich Poppitz}
\affiliation{Department of Physics, University of Toronto, Toronto, ON M4Y 3B6, Canada}
\author{Mithat \" Unsal}
\affiliation{SLAC and Physics Department, Stanford University, Stanford, CA 94025/94305, USA}
\preprint{}
\begin{abstract} 

We study the Eguchi-Kawai reduction in the strong-coupling domain of gauge theories via the gravity dual of ${\cal N}$$=$$4$ super-Yang-Mills on $\R^3\times \S^1$. We show that  $D$-branes geometrize volume independence   in the center-symmetric vacuum and give supergravity predictions for the range of validity of reduced large-$N$ models at strong coupling.

\end{abstract}

\maketitle
\twocolumngrid

\section{Introduction} 
\label{intro}
Gauge-gravity duality is a  powerful method 
to study strongly-coupled gauge dynamics. It  relates a weakly-coupled theory of  gravity to a lower dimensional large-$N$ gauge theory at strong coupling \cite{Maldacena:1997re,Gubser:1998bc,Witten:1998qj}.  The best-understood example  is the  AdS$_5$/CFT$_4$-correspondence between  
$\N$$=$$4$  supersymmetric Yang-Mills (SYM) theory on $\R^4$ and weakly coupled type-IIB supergravity on  AdS$_5 \times S_5$, where all available evidence suggests that the correspondence is exact, at least to leading order in $1/N$. 

  Another method   to study gauge dynamics in lattice and continuum formulations 
   is  the large-$N$  volume independence \cite{Eguchi:1982nm,Yaffe:1981vf,Bhanot:1982sh,GonzalezArroyo:1982hz,Kovtun:2007py,Unsal:2008ch}.  
  While much less appreciated than the AdS/CFT correspondence, large-$N$ volume independence is one of the few exact results in gauge theories.
  The  statement of the volume independence theorem is that
  large-$N$   non-abelian quantum gauge theories toroidally compactified on four-manifolds, 
 $M_4$$=$$\R^{4-k}\times ({ \S^1})^k$,    have properties
 that are independent of the $({ \S^1})^k$ compactification radii. More precisely, expectation values and connected correlators of single-trace operators are the same in the reduced and infinite-volume theories, to leading order in $1/N$---if the operators are neutral under the $(\Z_N)^k$ center symmetry and carry momenta  in the compact directions quantized in units of the inverse compactification radii. 
Volume independence holds provided  two basic quantum mechanical conditions are satisfied:
   {\it i.)}~translation symmetry is not spontaneously broken and
 {\it ii.)}~$(\Z_N)^k$ center-symmetry is not spontaneously broken. 
 
    In lattice-regularized gauge theories, where the lattice is reduced to a single site, this equivalence is known as ``large-$N$ reduction'' or  ``Eguchi-Kawai (EK)-reduction"  \cite{Eguchi:1982nm}.  The  necessary and sufficient conditions for the validity of volume independence  have been  known since the early 80's. However, 
  the first examples of gauge theories which satisfy them to arbitrarily small volumes  were found only recently \cite{Kovtun:2007py, Unsal:2008ch}. 
Because of this, there has been  a recent resurgence of interest in this  subject,  particularly in the lattice community---not only because small volume large-$N$ simulations are  more cost effective, but also for other reasons, such as lattice supersymmetry 
 \cite{Bringoltz:2009kb, Hietanen:2009ex, Ishii:2008ib,Ishiki:2009sg,Hanada:2009kz,  
 Hanada:2010kt}.  Furthermore, any gauge theory which satisfies volume independence  admits  a complementary volume-dependent domain, obtained by first fixing $N$ and taking the radii small,  where subtle non-perturbative aspects, such as the existence of a mass gap, 
 can be analyzed by semi-classical methods, see e.g., 
 \cite{Unsal:2008ch,Poppitz:2009uq,Meisinger:2009ne}.  The existence of a semi-classical domain is the main advantage of  studying the compactified theory, instead of the theory on $\R^4$. 
  For some  center-symmetric theories, 
 there is evidence suggesting that the  small radius domain  is the analytic continuation of the 
 large or infinite radius \cite{Unsal:2008ch} (also see \cite{Myers:2007vc, Pisarski:2006hz}),  
 and  the size of the circle times $N$ may be used as  an analytic expansion parameter.

   Volume independence holds  for arbitrary values of the coupling, including  the strong-coupling limit  of the gauge-gravity correspondence. It is thus interesting to examine the   consistency of the two correspondences; at the very least, this  provides a consistency check on their exactness. In this paper, we exhibit the simplest set-up where volume independence and AdS/CFT should hold simultaneously (see the concluding section for comments on related earlier work \cite{Furuuchi:2005eu}). We consider the gravity dual of strongly-coupled $\N$$=$$4$ SYM compactified on $\R^3 \times \S^1$ and study how volume independence arises.
We show that in the center-symmetric vacuum $D$-branes ``geometrize" volume independence, ensuring that 
 the expectation value of, e.g.,  a  Wilson loop in the uncompactified ($\R^3$) directions is independent of the $\S^1$ compactification radius, for arbitrary interquark separation and in accordance with the volume independence theorem.

  \section{Center-symmetry broken vacuum: volume dependence}
  
 The type-IIB background dual to $\N$$=$$4$ SYM compactified on $\R^3 \times \S^1$ of radius $R_0$ is:
    \begin{equation}
ds^2=  \frac{u^2}{R_3^2} (-dt^2 + \sum_{i=1}^{2} dx_i^2 + R_0^2 d \theta^2) +   \frac{ R_3^2}{u^2}  du^2 +  {R_3^2} d\Omega_5^2 . 
\label{D3-1}
\end{equation} 
This is  compactified $AdS^5$$\times$$\S^5$ of radius $R_3$$\sim$$\lambda^{1\over 4}$,  in local Poincare coordinates, expressed
in terms of the energy variable  $u\equiv r/l_s^2$. We use string units $l_s$$=$$1$ and denote  the 't Hooft coupling of the dual SYM theory by $\lambda$$\equiv$$g_{YM}^2 N$.  Compactification of a worldvolume direction of  AdS$_5$  leads to a conical singularity: as seen from (\ref{D3-1}), the proper radius of $\S^1$,  equal to $u R_0/R_3$, becomes of order the string scale at $u R_0 \sim \lambda^{1 \over 4}$.  The masses of Kaluza-Klein excitations and string winding modes become comparable, invalidating the supergravity approximation.
Thus, for energy scales $u R_0 < \lambda^{1\over 4}$ the non-singular gravity description is given by the T-dual (along the $x_3 = R_0 \theta$ direction)  type-IIA background of $N$ $D2$ branes  located on a dual circle of size $1/R_0$. 

\begin{figure}
\includegraphics[width=3.5in]{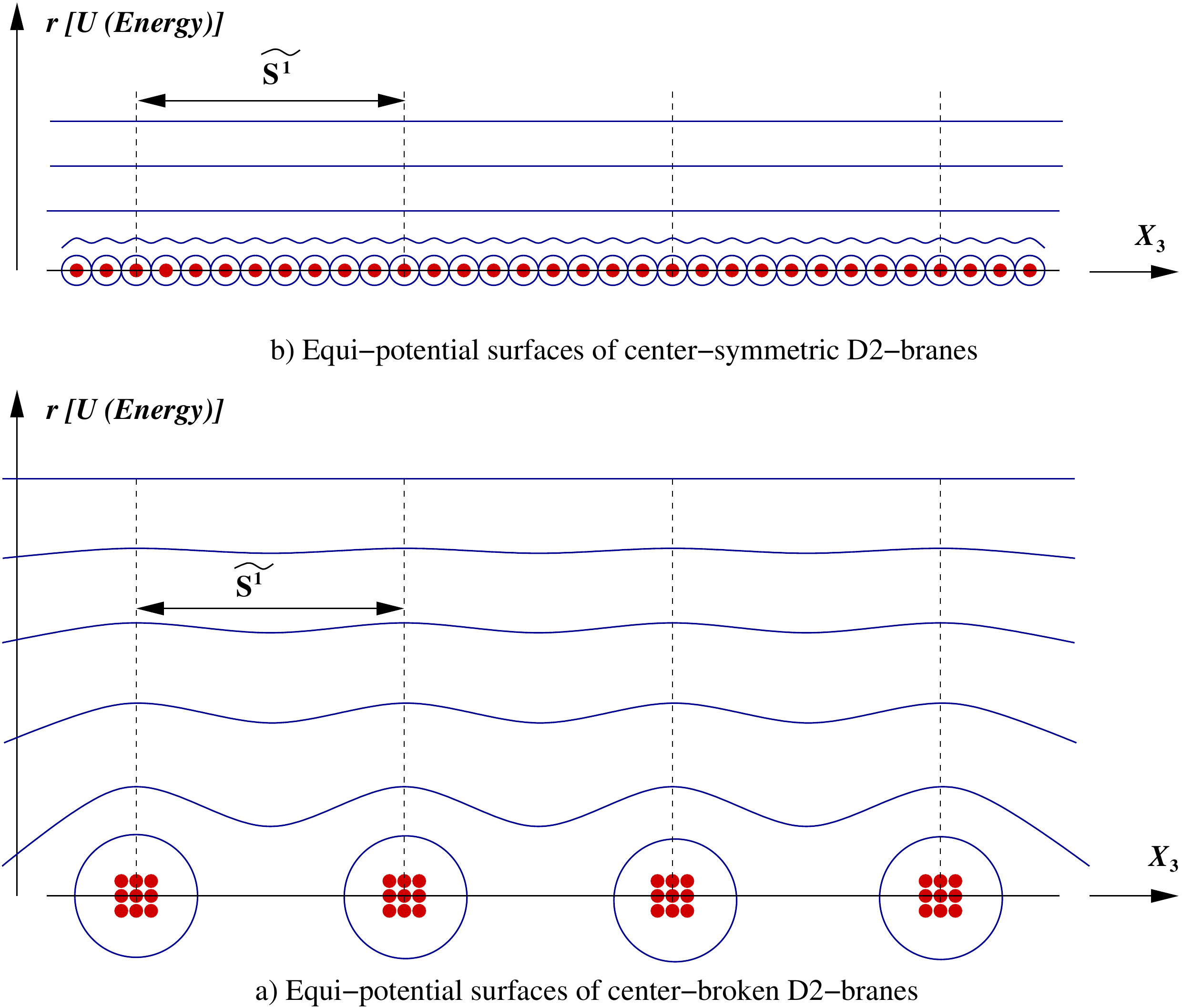}
\caption{Center-symmetric and center-broken D2-branes on the dual $\tilde{S}^1$ of radius $1/R_0$.}
\label{fig:equipotential}
\end{figure}

 The positions of the $D2$ branes on the dual circle correspond to the eigenvalues of the 
Wilson loop $\Omega \equiv {\rm exp}\left[ i \oint_{S^1} A \right]$ of the gauge field around the compact direction. Thus,  a vacuum 
 where all $N$ $D2$ branes are located at the same point on the dual circle breaks the center symmetry, as $\tr \Omega = N$.
The type-IIA gravity background corresponding to the center-broken vacuum is easy to determine by the method of images and knowledge of the 
 background of $N$ $D2$-branes in $\R^{1,9}$:
  \begin{equation}
  \label{D2-1}
ds^2= H_2(\bar r)^{-{1\over 2}}( -dt^2 + \sum_{i=1}^{2} dx_i^2) + H_2(\bar r)^{1\over 2} ( d \bar r^2 + \bar r^2  d\Omega_6^2 ).
\end{equation} 
Here $H_2(\bar r) = 6 \pi^2 g_s N/\bar r^5$ is a harmonic function in the seven dimensions transverse to the stack of $D2$ branes  and $g_s$ is the type-IIA string coupling. Instead of presenting detailed formulae (given in, e.g., \cite{Ponton:2000gi}),  for our purposes it suffices to only picture the brane arrangement. Taking $x_3$ as the compact direction, the metric of the center-broken BPS-brane configuration is determined by a harmonic function equal to the sum of the harmonic functions due to each stack of $N$ $D2$ branes separated a distance $1/R_0$ along $x_3$, as shown on Fig.~\ref{fig:equipotential}a. Each stack of branes creates an $1/\bar r^{5}$ ``potential," where $\bar r^2 = x_3^2 + r^2$ and $r$ $(= u)$ denotes the radial direction transverse to both the $D2$ branes and the compact direction. It is clear from the picture (and intuition from electrostatics) that when $u \gg 1/R_0$, the $x_3$-translational invariance of the background is recovered and that at $u = r\gg 1/R_0$ the harmonic function becomes  $\sim 1/r^{4}$,  identical to that of the corresponding stack of $D3$ branes (recall  that the type IIA coupling is $g_s = g_{YM}^2/R_0$). Thus the type IIA metric in the center-broken vacuum reads, for $u R_0 \gg 1$:
 \begin{equation}
ds^2= \frac{u^2}{R_3^2} ( -dt^2 + \sum_{i=1}^{2} dx_i^2)  + 
\frac{R_3^2}{R_0^2 u^2} d \theta^2  +   \frac{ R_3^2}{u^2}  du^2 +  {\hat R_3^2} d\Omega_5^2~,
\label{D2-2}
\end{equation} 
up to exponentially small corrections.
The metric (\ref{D2-2}) is the T-dual metric of  (\ref{D3-1}) (in the sense of \cite{Buscher:1987qj}) as evidenced by the fact that only the $d\theta^2$ terms are different (we do not show the relation between the type-IIA and type-IIB dilatons, which is trivial to obtain). It is also clear that  in the center-broken vacuum the type-IIA metric  will differ from (\ref{D2-2}) once $uR_0$ becomes of order unity or smaller.   

We conclude that in the center-broken vacuum, the   backgrounds (\ref{D2-2}) and (\ref{D3-1}) are equivalent  for $u R_0 \gg 1$, where the $x_3$ isometry is restored. Thus, for example, a calculation of the expectation value of a Wilson loop of size $R\times T$, positioned  in the $x_1$$-$$t$ plane of the noncompact $\R^3$ can be made via (\ref{D2-2}) so long as $R \ll R_0$---so that the string worldsheet  only probes the bulk geometry in the  $u R_0 \gg 1$ region, close to the ``UV-brane" (recall the ``energy-distance" relation, $u_{*} R$$\sim$$1$, for the minimum value $u_*$ of $u$ probed by a Wilson loop of size $R$ \cite{Maldacena:1998im}). 
Thus, Wilson loops of interquark separation $R \ll R_0$ are unaffected by the compactification, as one would naively expect. However, the worldsheet relevant for Wilson loops with $R \gg R_0$  probes the bulk geometry further away from the UV, as  now $u_* R_0 \ll 1$, a region, where (\ref{D2-2})   receives 
corrections due to the compactification.

Hence, in the center-symmetry breaking vacuum, the Wilson loop (and other correlators) exhibit volume dependence. This is consistent with expectations from compactified field theory that the  Wilson loop with inter-quark separation $R \gg R_0$ should be sensitive to the $\R^3 \times \S^1$ compactification.
In fact, a dual gravity analysis  \cite{Itzhaki:1998dd} of the Wilson loop in the compactified $\N=4$ SYM theory  shows that the behavior of the quark-antiquark potential changes from $1/R$, at short distances $R \ll R_0$, 
to $1/R^{2\over 3}$ in an intermediate $D2$-brane region, and back to $1/R$ in the far-infrared $M2$-brane region (the latter describes the three dimensional 16 supercharge CFT that $\N =4$ SYM flows to upon a center-symmetry breaking compactification \cite{Seiberg:1997ax}).

   \section{Center-symmetric  vacuum: volume independence}
   
 Consider now the center-symmetric vacuum of the $\N = 4$ SYM theory on $\R^3 \times \S^1$. 
  According to EK reduction, appropriate observables should now exhibit $\S^1$-size independence. 
   
   In the type-IIA picture, a center-symmetric vacuum corresponds to a configuration of  $N$ $D2$ branes distributed equidistantly on the dual circle---since their positions on the dual $\S^1$ correspond to the eigenvalues of $\Omega$,   now clearly $\tr \Omega^k = 0$, for all $k\neq 0{\rm (mod)} N$. The metric dual to the center-symmetric $D2$-brane configuration can similarly be computed using the method of images. The difference is that now single $D2$-branes are spaced a distance $1/(N R_0)$ apart along the compact $x_3$ direction, as shown on Fig.~\ref{fig:equipotential}b. The harmonic function determining the background is, again, the sum of the $1/\bar{r}^5$ ``potentials" of the individual $D2$-branes (with $\bar r^2 = x_3^2 + r^2$, as before), resulting in the $D2$-brane harmonic function which determines the metric, as in (\ref{D2-1}):
\begin{eqnarray}
 H_2^{\rm sym}(r, x_3) =  
\sum_{n=- \infty}^{\infty} \sum_{k=1}^N \frac{  6 \pi^2 g_s }{\left[ r^2 +  (x_3 -  
 \frac{2 \pi}{R_0N}k  - \frac{2 \pi n}{R_0})^2 \right]^{5 \over 2}} . ~
 \label{H2sym}
 \end{eqnarray}
By Poisson resummation,  
(\ref{H2sym}) takes the form:
\begin{eqnarray}
{R_3^4 \over u^4} \left[1+ \sum_{m=1}^{\infty} \left( m u NR_0 \right)^2 K_2( m u NR_0) \cos(m x_3 NR_0) \right],~
 \label{H2sym2}
 \end{eqnarray}
where  $K_2$ is the modified Bessel function.  

The crucial difference 
 with respect to  the center-broken vacuum discussed in \cite{Ponton:2000gi} is the appearance of  a factor of $N$ in the correction term in (\ref{H2sym2}). 
 Hence,  the $x_3$-isometry is now recovered for much smaller values of $r$ ($=u$, the energy scale). It is clear (from (\ref{H2sym2}) or from electrostatics) that now the condition for isometry restoration is $u N R_0 \gg 1$, instead of $u R_0 \gg 1$ in the center-broken vacuum. 
 Thus the background dual to the center-symmetric vacuum is also given by (\ref{D2-2}), but is now valid for $uNR_0 \gg 1$.   We note that while near each individual $D2$-brane the supergravity approximation is not to be trusted (because large curvatures occur and physics is described by an IR free abelian theory), no large curvatures appear in 
the background (\ref{D2-2}), (\ref{H2sym2}) at $u \gg 1/(NR_0)$, i.e., at any finite distance away from the D2-branes (similar backgrounds are also considered in \cite{Lin:2005nh}).

 The fact that the center-symmetric vacuum is described by  (\ref{D2-2}) for any $u \gg 1/(NR_0)$ immediately implies  that  a Wilson loop of any size (strictly speaking of size $R \ll N R_0$) will be insensitive to the compactification. Thus, the potential between two static quarks exhibits the behavior characteristic of the four dimensional $\N = 4$ CFT, $V(R) \sim \lambda^{1\over 2}/R$,  at all scales, despite the fact that one dimension is compactified and conformal symmetry of the background is 
 explicitly broken. 
 
 In the language of non-perturbative orbifold equivalences
 \cite{Kovtun:2007py},  the neutral sector observables in the compactified  ``daughter" theory enjoy the conformal symmetry of its ``parent"  theory on $\R^4$, at leading order in $N$.
 However, it should also 
 be noted that  the daughter theory also possesses a non-neutral sector aware of the compactification radius. The main point is that for neutral-sector  observables,  the space may be viewed as having an effective size  $R_{\rm eff} =R_0  N$ and thus $N=\infty $  is a decompactification limit. 
 
  It is   clear that other quantities will also exhibit volume independence---for example, correlation functions of single-trace operators that only carry momentum in the noncompact directions will also be insensitive to the compactification, as required by EK reduction. EK reduction also requires that expectation values of Wilson loops extending also in $\S^1$, but not winding around the compact direction (i.e., center-symmetry neutral ones),  exhibit volume independence; however, explicitly verifying their volume independence in the gravity dual appears more challenging to us than for the observables we consider.

 \begin{figure}
\includegraphics[width=3.5in]{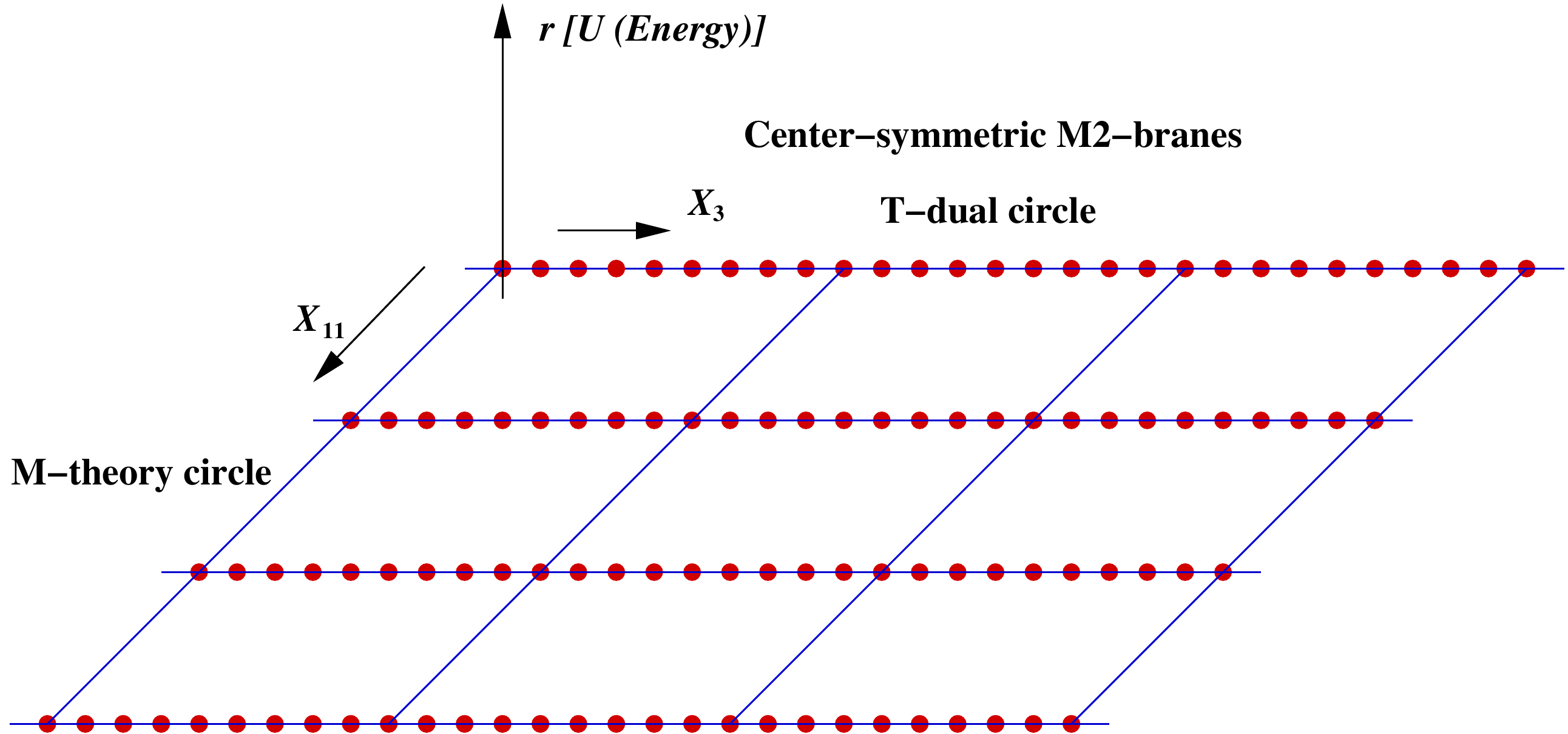}
\caption{Center-symmetric $M2$-branes.}
\label{fig2}
\end{figure}

Finally, we briefly note a   slight refinement of the condition for volume independence inferred from the gravity dual. In our  discussion above, we did not consider the behavior of the type-IIA dilaton. In fact, examining its behavior shows that the effective string coupling becomes large when $u N R_0\sim \lambda^{5 \over 4}$---thus,  the region of validity of volume independence  that can be inferred from the type-IIA dual of the center-symmetric vacuum would be  $u N R_0 \gg \lambda^{5 \over 4}$, instead of simply $u N R_0 \gg 1$.
However, the type-IIA description can be uplifted to eleven dimensional supergravity (M-theory) \cite{Itzhaki:1998dd}. The size of the eleventh direction, parameterized by $x_{11}$, is related to the type-II couplings as $R_{11} = g_s = \lambda/(N R_0)$ and the center-symmetric $D2$ brane configuration is replaced by a similar configuration of $M2$-branes located a distance $1/(N R_0)$ apart along $x_3$, as shown on Fig.~\ref{fig2}. For $u \gg R_{11}= {\lambda \over N R_0} \gg 1/(N R_0)$ the $M2$-brane background can be written in a form:
\begin{eqnarray}
ds_{11} &&= e^{-2\Phi/3} ds_{10}^2  +  e^{4\Phi/3} dx_{11}^2 ~,
\end{eqnarray}
where $ds_{10}^2$ is the  type-IIA metric (\ref{D2-2}) and $\Phi$ is the type-IIA dilaton. As this background is dual to (\ref{D2-2}), it follows that the regime of volume independence is $u N R_0 \gg \lambda$, improving on the type-IIA bound $u N R_0 \gg \lambda^{5 \over 4}$. A bound on the validity of volume independence  of the form $u N R_0$$\gg$${\rm max} (1, \lambda)$ can also be inferred from field theory considerations and is  consistent  with the strong-coupling bound from supergravity obtained here (see \cite{Unsal:2010qh} for a field-theory analysis of ${\cal{N}}$$=$$4$ SYM in the volume-independence context). 
At asymptotically low energies $u N R_0 \ll 1$, eleven dimensional supergravity breaks down for center-symmetric M2-branes, consistent with the free abelian  long-distance dynamics, unlike the coincident M2-branes case where the IR-physics is non-abelian and superconformal.

\section{Conclusions}

We considered the simplest case where EK reduction of a four-dimensional gauge theory is valid simultaneously with gauge-gravity duality.
Our considerations indicate that $\N=4$ SYM compactified on $\R^3 \times \S^1$ indeed exhibits volume independence in the center-symmetry preserving vacuum. The gravity dual of four dimensional $\N=4$ SYM gives   the first explicitly solvable realization of volume independence above two dimensions (where EK reduction is manifest in the large-$N$ limit of the exactly solvable pure YM lattice theory  \cite{Gross:1980he}).  Our findings can also be viewed as providing a check on the weakest form of the AdS/CFT correspondence. 

It would be interesting to consider how EK reduction works when more than one dimension is compactified, especially with regard of how center-symmetry preservation is reflected in the brane  and gravity set-ups. The $\R^3 \times \S^1$ case is special in this respect, as one is  free to  choose a classical center-symmetric vacuum state, not washed away by quantum fluctuations which become strong as more dimensions are compactified. We note that ref.~\cite{Furuuchi:2005eu}  previously considered large-$N$ reductions in the holographic picture with all dimensions  compactified, but   the matching of observables   and the question of fluctuations raised above were not studied.

It may also be interesting to study volume independence for confining gauge theories  
with known  gravity duals,   as well as by exploiting the analogy between 
the $1/N$ and  genus expansion in  gauge and string theories. For example,   \cite{Atick:1988si} showed that free energy of YM theory receives contributions only from Riemann surfaces of genus $\geq 1$ in the confined phase $[O(N^0)]$, but it receives a contribution from genus zero in the deconfined phase  
$[O(N^2)]$. This is nothing but the   temperature independence of confined phase and temperature dependence of the  deconfined phase, to leading order in 
$N$. In this context, for example, the double-trace deformation stabilizing the theory to the confined phase   \cite{Unsal:2008ch, Meisinger:2009ne} may have a  stringy interpretation in terms of analytic continuation of the winding-number unbroken phase of \cite{Polchinski:1991tw} to small radii.

\section{Acknowledgements}
  We are indebted to  J. Maldacena,  O. Aharony, and L. Yaffe  for their crucial help. 
     We also thank 
  B.  Burrington, M. Hanada, A.  Karch, H. Neuberger, A. Peet,  A. Rajaraman, S. Shenker, and E.  Witten  for useful discussions. 
     M.\"U.  thanks the Weizmann Institute of Science,   where portions of this work was done,
      for hospitality.       
 This work was supported by the U.S.\ Department of Energy Grant DE-AC02-76SF00515 and by the National Science and Engineering Council of Canada (NSERC).


\end{document}